\documentclass[twocolumn,aps,floatfix,superscriptaddress]{revtex4}

\usepackage{graphicx,xcolor}
\usepackage{hyperref}
\usepackage{bm}

\usepackage{epsf}
\usepackage{amssymb}
\usepackage{wrapfig}
\usepackage{amsmath,amsfonts}

\newcommand{\la}{\label}
\newcommand{\bea}{\begin{eqnarray}}
\newcommand{\eea}{\end{eqnarray}}
\newcommand{\beq}{\begin{equation}}
\newcommand{\eeq}{\end{equation}}
\newcommand{\be}{\begin{equation}}
\newcommand{\ee}{\end{equation}}
\newcommand{\ii}{{\rm{i}}}
\newcommand{\dd}{{\rm{d}}}

\newcommand{\p}{\partial}

\begin{document}
\title{Weak solution for the Hele-Shaw problem: viscous  shocks and singularities}

\author{S-Y. Lee}
\affiliation{Mathematics 253-37, Caltech, Pasadena, CA 91125, USA}
\author{R.Teodorescu}
\affiliation{Mathematics Department, Univ. of South Florida, 4202 E. Fowler Ave, Tampa FL 33620, USA}
\author{P. Wiegmann}
\affiliation{The James Franck  Institute, University of Chicago, 5640 S. Ellis Ave, Chicago IL 60637, USA}

\begin{abstract}

In Hele-Shaw flows a boundary of a viscous fluid develops unstable fingering patterns. At vanishing surface tension, fingers  evolve to cusp-like singularities  preventing a smooth flow.  We show that the Hele-Shaw  problem admits a {\it weak solution} where  a singularity triggers  {\it viscous shocks}. Shocks form a growing, branching tree of a line distribution of vorticity where pressure  has a finite discontinuity.  A condition that the flow remains curl-free at a macroscale uniquely determines the shock graph structure. We present a self-similar solution describing shocks emerging from a generic (2,3)-cusp singularity -- an elementary branching event of a branching shock graph.

\end{abstract}

\maketitle

\paragraph*{1. Introduction} 

Hele-Shaw flow describes a 2D viscous incompressible fluid with a free boundary at  low Reynolds numbers. The fluid is either  sucked out from a drain or driven to a drain by another, inviscid, incompressible liquid \cite{Lamb}. Darcy law governs the viscous flow: 
\be \la{darcy}
\mathbf{ j} = -K \mathbf{\nabla} p,\quad \mathbf{ j}=\rho_0\mathbf{v}.
\ee
If  density $\rho_0$ and  hydraulic conductivity $K$ are  constants, then pressure in the incompressible fluid $p$  is a harmonic function, $\Delta p=0$. At a drain (set at infinity), the fluid disappears with  a  constant  flux $Q=\oint_\infty  \mathbf{ j}\times\dd \mathbf{ \ell}$.

At vanishing surface tension (the case we consider), pressure is a constant 
along a boundary. Thus, in the fluid, pressure is a solution of the Dirichlet boundary problem. 
The boundary itself evolves in time.

A compact form of the law involves only a boundary:  normal velocity  of a line element of the boundary 
$\dd\ell$  
is proportional to its  harmonic measure 
${\rm v}\dd \ell\sim \dd\mu$ 
(Harmonic measure  is a distribution of a Brownian excursion (emanating  from a drain) as it hits the boundary $\dd\mu=|\nabla  p|\,\dd\ell$). 

In this form, the Darcy law goes far  beyond applications to fluid dynamics \cite{Couder}. It is closely related to a wide  class of 2D growth and solidification processes such as  DLA \cite{DLA81}, flows in granular media \cite{Jag-Nag}, visco-elastic flow \cite{Maher}, evaporative patterns of  fluid monolayers \cite{Lipson}, etc.  

Common patterns observed in these flows are
characterized by intricate viscous fingering  instabilities  \cite{ST}. 

As such the problem is ill-defined:  at a finite, {\emph {critical time}} fingers develop cusp singularities (points of infinte curvature) \cite{bs84, Howison85,Hohlov-Howison94}, when the Darcy law stops making sense.  Nevertheless, in many important physical problems such as flows in granular media and solidification, which at large scales are described by Darcy law,  flows are not limited and go over a singularity. Resolving finite-time  singularities in a physical manner, and a description of a flow beyond singularities  is a major long-standing  problem in the field.

An origin of singularities is the approximation of all physical parameters which could  stabilize a flow at a microscale being set to zero. While this is a valid  approximation at a smooth flow, at  a singularity  perturbations are singular, and  an order of limits when different physical parameters are brought to zero is essential.  One  perturbation regularizing a flow  is  surface tension. However, interesting patterns are observed in processes where surface tension is either small, or does not exist at all, like in solidification and granular materials. This is the regime we are interested in. We assume that the  surface  tension (if any)  must be set to zero first, before other parameters such as  compressibility, are  set to zero as well.  

In this Letter we present
what we believe might be the solution to the problem of finite-time singularities in the Hele-Shaw problem.

We impose  the incompressibility and curl-free conditions  at a large scale, but  relax them  at a microscale, setting surface tension to zero in the first  place. Under this setting the Hele-Shaw problem admits a {\it weak solution}: 
a  singular cusp-like finger  triggers {\it shocks} -- lines
of   discontinuity for pressure. 

Shocks propagate through the fluid, forming a growing,  {\emph{branching tree}}.
In this Letter we analyze a  generic cusp-singularity giving rise  of  an elementary branching event of what will become a complicated degree-two tree of shocks.

The solution describing local  origin of viscous  shocks or any further branching event of already existing shocks pattern is  self-similar and does not depend on the details of the flow.
  
A distinct feature of the Hele-Shaw flow is integrability \cite{Galin, PK, Richardson72,us1, us3, us4, us5, us6}.
Our weak solution is the only regularization of singularities which preserves the flow integrability.

\begin{figure}
\includegraphics[width=0.30\textwidth]{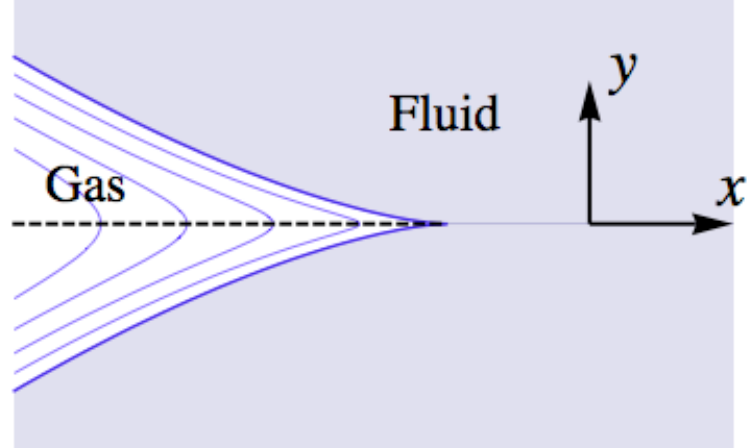}
\caption{A growing finger in local Cartesian coordinates. The dashed line is the branch cut of the height function.}
\label{finger}
\end{figure}

A comment is in order: the most common  appearance of shocks in hydrodynamics are ultrasonic flows in compressible fluids at vanishing viscosity. Formally they are caused by inertial terms in hydrodynamic equations. Shocks in Hele-Shaw flow considered here are of a different nature. They occur at vanishing Reynolds numbers when inertial terms are negligible and viscosity is high. We refer to them as {\it viscous shocks}.
 \newline 
 \indent \emph{2. Analytic and integrated forms of the Darcy Law }  
 We recall a description of a viscous finger in terms of  a  \emph{height} function \cite{us4}. In Cartesian coordinates aligned with a finger axis (FIG.\ref{finger}), a  finger  is given by  a graph $y(x)$.  Let $X=x+iy$ be a complex coordinate of the fluid. The height function   $Y(X,t)$ defined in the fluid is an analytic function whose boundary value on the fluid boundary  is  $Y(X)_{X=x}=y(x)$.
Its  discontinuity across the branch cut   is 
$\mbox{disc}\,\overline{ Y}|_{X=x}=2y(x)$. 
In terms of the height function the Darcy law reads:
   \be\la{darcy2}
\rho_0  \dot Y= -\p_X\phi,
  \ee 
  where the analytic function $\phi=\psi+\ii p$ is a complex potential of the flow, and $\psi$ is a  stream function.

An important physical characteristic is the (time-depedent) capacity of the flow. It is  defined as  $C(t)=\frac{1}{2\pi \ii}\oint_\infty \phi\,\dd\phi$, where the integral goes around a drain. According to  Darcy law, the power required to drain the
 fluid $N(t) = \frac{1}{2\pi K }\oint_\infty p\, (\mathbf{j}\times\dd \mathbf{\ell})$  yields  capacity $N(t)=-C$.

{\it Integrated form of Darcy law:}
The integral $\Omega(X)= -\ii \int_e^X\rho_0 Y\dd X$ 
  gives yet another form of the Darcy law:
 \be\la{darcy3}
 \dot \Omega = \ii \phi\,=-p+\ii\psi. \ee
Let us  integrate (\ref{darcy2})  over a cycle $B$ in the fluid:
 \bea \la{im}  \frac{\dd}{\dd t}{\rm{Im}}\, \oint_{B}   \dd \Omega
&=&\!\!\oint _B \mathbf{j\times d\ell}
= \oint_B \dd\psi,\\
\la{re}  \frac{\dd}{\dd t} {\rm{Re}}\,
\oint_{ B}    \dd \Omega &=& - \oint _B  \,\mathbf{
j\cdot d\ell} 
=  - \oint_B \dd p=0.
\eea 
The imaginary part measures a flux of fluid  through the cycle, the real part measures circulation along the cycle.  The latter  vanishes. 
At infinity, $\ii \oint_\infty \dd\Omega=Q t$  represents the mass of fluid drained up to time $t$.

 \paragraph*{3. Singularities of the Hele-Shaw flow} The pressure gradient is highest where the boundary curvature is large. Thus by \eqref{darcy}, growth velocity is largest (and increasing) at finger tips. It  diverges at a  critical time; the finger then becomes a  cusp of type $(2, 2l+1)$:   $y^2\sim x^{2l+1}$ \cite{bs84, Howison85, Howison86, 
Hohlov-Howison94,us4},  FIG.~\ref{finger}. 

Cusps of  types $(2,4k-1)$ and $(2,4k+1)$ evolve  differently.  In \cite{us5} it was shown that for the type $(2, 4k+1)$,  a new droplet of inviscid fluid emerges next to the finger tip, before it  evolves into a cusp. The fluid  becomes multiply-connected, but evolution continues  smoothly.
 
 No continuous evolution is possible for\!  $(2,4k\!-\!\!1)$\! cusps, including the most generic cusp (2,3) 
 --  subject of this paper. 

\paragraph*{4. Hydrodynamics of a critical finger} 
  A  finger approaching the (2,3)-cusp is especially simple \cite{us3}. 
Fixing a scale and the origin, it is given by:
 \be\la{e}
- Y^2=4X^3-g_2X-g_3=
 4(X-e_3)(X-e_2)(X-e_1),
  \ee
  where $g_2$ and $g_3$ are real time dependent coefficients. One of  the branch points,  $e_3$, may be chosen real. It is the tip of a finger. The other two  are conjugated $e_1=\bar e_2$.  
  The coefficient $g_2=4(|e_1|^2-e_3^2)$ is determined by the drain rate,  $Q $. We  set the rate  that $g_2=-12t$, and  count time from a cusp event:  flow is smooth at $t<0$. 
  
  In this normalization, capacity equals $C=-\dot g_2 $.  Well before a critical time, $e_3<0$.  Then at $t<0$, $g_2>0$ and $\mbox{Re}\,e_1>0$: the two  branch points   are located in the fluid. This  breaks the conditions of incompressibility, unless the roots coincide to a real  double point:  $e_1=e_2$.    Thus, $g_3=8(-t)^{3/2}$. The height function  at $t <0$ is a degenerate elliptic curve \cite{us3}:
 \be\la{133}
Y^2=-4 \left(X-e(t)\right)\left(X+e(t)/2\right)^2,\:\:\: e(t)=-2\sqrt{-t}.
\ee
 A finger becomes a cusp when the branch point $e$ and double point $-e/2$ merge to a triple point. 
An important property of the critical finger (\ref{133})  is that it is  self-similar:
\be\la{s}
Y(X,t)=|t|^{3/4}Y\left(|t|^{-1/2}X,1\right).
\ee
  
   \paragraph*{5. Weak form of the Hele-Shaw flow} 
  
  Once  the flow reaches a  cusp singularity, it is no longer governed by the differential form of the Darcy law.
This situation is typical for conservation laws of hyperbolic type 
$\p_t {{u}} + \partial_z f(u) = 0$.
There as well, smooth initial data develop into a shock at a finite time.  
Shocks occur to ill-defined conservation equations  which arise as approximations of a well-defined problem. Adding a deformation through  terms with higher gradients  prevents formation of singularities. However, a  smooth solution of a deformed equation may become   discontinuous 
as a deformation is removed. It is  then called a {\it weak solution} \cite{Evans}.  Validity of the differential equation  on both sides of a shock determines a traveling  velocity of a front  $
V=\frac{{\rm disc}\,f}{{\rm disc}\, u}
$- the Rankine-Hugoniot condition 
\cite{LL}.

Often,  physical principles determine dynamics of shocks without specific knowledge  of the deformation used,
 such that different deformations lead to the same weak solution.
The best known example is the Maxwell rule determining the  position of Burgers-type shock fronts \cite{LL}.

 Darcy law  is a conservation law of  hyperbolic type, and we adopt the same strategy. We will be looking for a weak  solution of the Hele-Shaw  problem when the Darcy law is applied everywhere in the fluid except on a moving, growing and branching graph
$\Gamma(t)$  of viscous shocks  (or cracks), where pressure  suffers a finite discontinuity.

A  few  natural physical principles   guide toward a unique weak  solution. We give three equivalent formulations. 

The first formulation is in terms of the height function, treated as a complex vector:
\begin{enumerate}
\item [-] The canonically  oriented (anti-clockwise) discontinuity of the height function's complex conjugate is parallel to  the shock line directed such that: 
\be\la{jump1} 
  {\rho_0}\,{\rm{disc}}\; \overline Y|_\Gamma=-2\sigma{\bm\ell},\quad \sigma=\rho_0 |Y|>0\ ,
\ee 
where ${\bm\ell}$ is a unit vector canonically oriented along the shock line.
\end{enumerate}
An equivalent  invariant formulation  is in terms of the generating function (\ref{darcy3}):
\begin{itemize}
\item [-]  Discontinuity of $\Omega$ on shocks is imaginary; ${\rm Re}\,{\rm{disc}}\,\Omega$  is {\it increasing} away from both sides of  a shock,
\be\la{A1}
{\rm Re}\,{\rm{disc}}\,\Omega|_\Gamma=0, \quad {\rm Re}\,\Omega|_{X\to \Gamma}>0.
\ee
\end{itemize}

The first  condition in (\ref{A1})  is equivalently written as $\mbox{Re}\,\ii\oint Y\dd X=0$ for all cycles. Curves of  this kind are  called  {Krichever-Boutroux} curves.   They   appeared in studies of Whitham averaging of non-linear waves \cite{Krichev-red} and asymptotes 
 of orthogonal polynomials  \cite{Bertola-Mo}.  Neither appearance is coincidental. 

The third formulation is in hydrodynamics terms.


\paragraph*{6. Hydrodynamics of  viscous shocks}  We require  that the fluid is irrotational at macroscale, but allow  vorticity at a microscale -- i.e. at a scale set by a vanishing regularizing parameter.

The first conditions of  (\ref{jump1},\ref{A1}) insure that  the fluid remains {\it curl-free} despite of shocks. Let us chose a contour $B$ including a portion of a shock. Then   (\ref{jump1},\ref{A1}) mean that the integral (\ref{re}) still vanishes despite of discontinuities of  $\Omega$ and pressure. 

 Let us study this condition in detail. The time derivative  has  two
contributions: one from  evolution of the  height function $Y(X,t)$ at a fixed $X$, another
from motion of shocks. Denote the shock front velocity by ${\bf V}_\perp$ (normal to the front, 
directed along the vector $\bf n=-\ii{\bm \ell} $
). Then the  time derivative in (\ref{re}) reads
$
{\rm{Im }} \, [{\rm{disc }} \, \dot Y \,{\bm \ell} +\nabla_\parallel ({\rm{disc }} \, Y \cdot {V}_\perp) ]_{ \Gamma}
=0,
$
where $\nabla_\parallel$ represents the derivative along the direction tangent to the front along the vector $\bm\ell$. 
The first term is a jump of the velocity of the fluid parallel to a shock  $ - {\rm{disc}} \,\, v_{\parallel} $.
The second term is purely 
real. It equals $\nabla_\parallel \left(\sigma {\bf V}_\perp\right) $. 
Together, they yield to the condition:
\be\la{C}
\nabla_\parallel{J_\perp} + \,{\rm{disc}} \,\, j_{\parallel} = 0,\quad {J_\perp}= \sigma  { {\bf V}_\perp},\quad j_\parallel=\rho_0 \,v_\parallel.
\ee
The first term in this equation represents the transport of mass by  a shock (normal to the shock), while the
second is a circulation of the surrounding fluid flow. They compensate each other. 

This condition, derived solely from the requirement that $\sigma $ is real, suggests to interpret viscous  shocks as a single layer of microscale vortices with a line density  $\sigma $ and fixed orientation.
Therefore, the weak solution describes a fluid with zero vorticity at a macroscale: vorticity concentrated in shocks  is compensated by  the circulation of fluid  around shocks.

Using Darcy law we  replace the fluid velocity in (\ref{C}) by
$-\nabla_{\parallel} p$, and integrate (\ref{C}) along the cut. We obtain the  Rankine-Hugoniot condition:
\be\la{rh1}
 \sigma  {\bf {V_\perp}}=({\rm {disc}}\;p)\;{\bf n}\ .
 \ee

The  second set of conditions in (\ref{jump1},\ref{A1}) ($\sigma>0$)  imply that shocks move toward higher pressure, i.e, represent a deficit of  fluid - cracks. This  follows from the curl-free condition and is  consistent with an 
assumption that  at  infinity (at a remote part of a  finger,  and at a drain), the flow is  not affected by shocks.  

 Consider the integral around the drain $\ii\oint_\infty \dd\Omega=Q t$.  Before the transition,  a  contour of integration can be smoothly deformed to the fluid boundary. It  yields the total mass ${2}\rho_0\oint y(x)\dd x=\rho_0\int\!\!\!\int \dd y\dd x$. After the transition, the integral acquires  a contribution from  shocks: $-\oint_\Gamma \sigma\dd\ell$.  Viscous shocks can be seen as 
 a single layer mass deficit: density of the fluid is no longer constant, $\rho(X)=\rho_0-\delta_\Gamma(X) \sigma(X)$  (here $\delta_\Gamma$ is the delta-function on  shocks).
 
 The Rankine-Hugoniot condition (\ref{rh1}), curl-free condition (\ref{jump1})  and  the differential  form of the Darcy equation (\ref{darcy2}) combined give the {\it weak form of the Darcy law}.  
 Remarkably, these conditions uniquely determine a shock pattern. 
 Shock graphs with one and two branching  generations are represented in FIGs.~2,3. We will show how  they work by an  analysis of the (2,3)-cusp .


\paragraph* {7. Self-similar weak solution: beyond the (2,3)-cusp} 
In the remaining part of the Letter  we  briefly  describe a  solution to the most generic singularity, (2,3).

Before becoming cusps, fingers are described by (\ref{133}).  The height function is a self-similar  degenerate elliptic curve.
Two branch points located in the fluid coincide to a double point.
At a critical time,  the tip meets the double point. 
After the critical time, we look for a weak solution, allowing pressure to have finite discontinuities on some curves. We assume that infinity is not affected by the transition, such that the height function is still an elliptic curve (\ref{e}) with $g_2=-{12}t$, but now $g_2<0$. We have only to find  $g_3$ from the condition (\ref{A1}).   Solution is again self-similar, $g_3(t)=g_3(1) t^{3/2}$. Once $g_3(1)\neq 8$, the curve is not degenerate anymore.
The double point splits into two branch points, $e_1\neq e_2$. They appear in the fluid as endpoints of shocks.

\begin{figure}[htbp]
\begin{center}
\includegraphics[width=4.25cm]{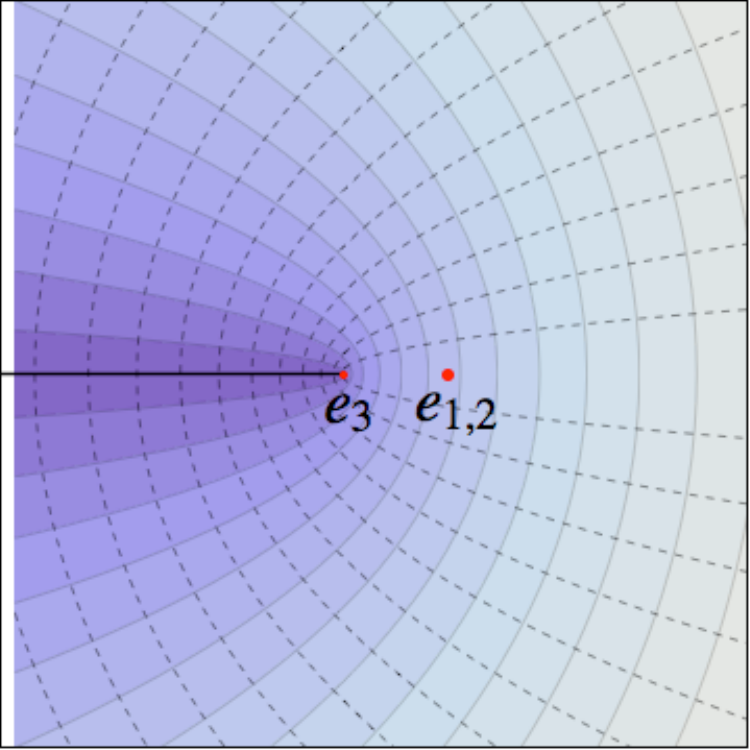}
\includegraphics[width=4.25cm]{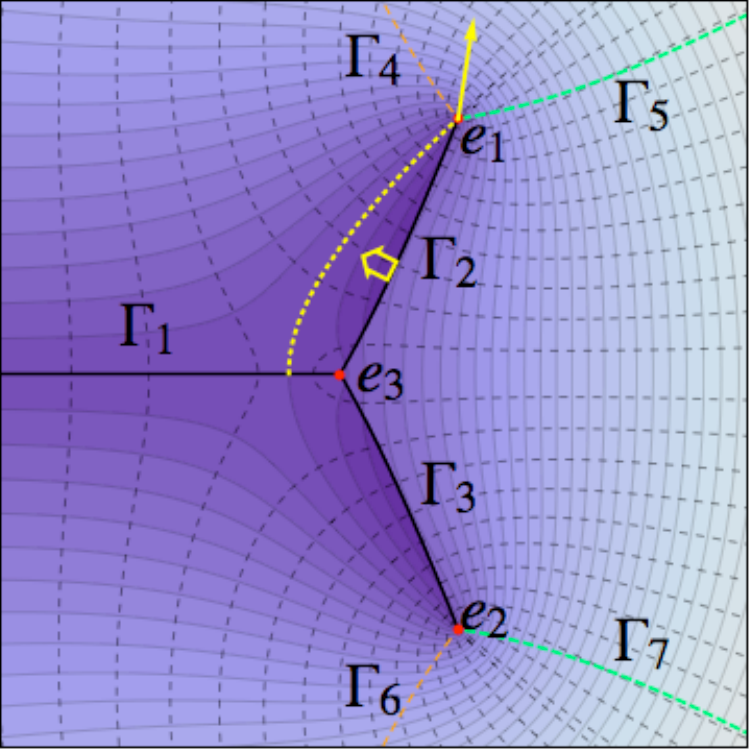}
\caption{The equi-pressure lines and flow  lines (dashed) before the  
transition (left), and after the transition  (right).    Pressure  gets larger as the shade gets  
darker.   Before the transition 
$e_2=e_1$ is the double point. The  
thick lines (on the right panel) connecting the branch points  are  
the shocks.  The orange and green dashed lines  are not admissible  the level lines of ${\rm Re}\,\Omega$.
The fluid flows to the lighter  region toward low pressure. Shocks moves to toward the darker region (higher pressure).
A bright dotted line
 is the zero-pressure line. The arrows are moving directions of the shock and the branch point. On the left of the zero-pressure line the finger  expands, on the right the finger retreats.}\label{beforeafter}
\end{center}
\end{figure}

 Since  branch points are simple, all quantities have opposite signes on opposite sides of  shocks:  $\mbox{disc}\,\Omega=2\Omega|_\Gamma$. Then condition (\ref{A1}) means that $\mbox{Re}\, \Omega$ vanishes on shocks. Since the  endpoints belong to the fluid, by virtue of (\ref{darcy3}), pressure also vanishes. The branch point $e_3$ is a tip of a finger where pressure obviously vanishes.  
 This is the  governing condition:
 \be\la{p}
  p(e_{1,2})={\rm Im}\,\phi(e_{1,2})=0.
  \ee
 The scaling property  (\ref{s}) is sufficient to 
 express the potential $\phi$ and $\Omega$ through elliptic integrals:
 \bea \la{phi}
 &&\phi(X)=6\int^{X}_{e_3}\left(X+\frac{3}{2}\frac{g_3}{g_2}\right )\frac{dX}{Y(X)},\\
 &&\Omega(X)=-{\rm i}\int_{e_3}^X Y\,dX=-\frac{2\rm i}{5}(XY-2t\phi).\la{Omega}
\eea
 The governing equation (\ref{p}) is then expressed through complete elliptic integrals  $K$ and $E$  \cite{Gamba}:
 \begin{equation}\label{EK}
(16m^2-16m+1)E(m)=(8m^2-9m+1)K(m)\ ,
\end{equation}
 where  $m=\frac{1}{2}+\frac{3}{2}\frac{1}{\sqrt{9+4h^2}}$ and $\frac 32\frac{g_3}{g_2t^{1/2} }=3\sqrt\frac{3}{4h^2-3}\frac{4h^2+1}{4h^2-3}$.
 Solution of (\ref{EK}) is transcendental \cite{Gamba}:
 \bea\la{nn}
&&-\frac{e_{1,2}}{e_3}=\frac 12\pm ih,\quad h=3.24638225374\dots
 \eea
 It determines the moving ends of shocks.
 
 \paragraph*{8. Discontinuous change of capacity}
 The genus transition is summarized by an abrupt change of $g_3|_{t=\pm 1}$. {We express this fact in normalization-independent terms using time derivative of the capacity}.
The ratio    
\begin{equation}
\eta=\lim_{t\rightarrow 0}\frac{\dot { C}_{t>0}}{\dot { C}_{t<0}}=\frac{3}{2}\frac{g_3}{g_2}t^{-1/2}
 =0.91522030388\dots \label{conjecture}
\end{equation}
 {is a unique {\it universal} number  describing the transition.}

\paragraph*{9. Flow and  shocks.}
Shocks are anti-Stokes lines of $\Omega$, i.e., zero-level lines of ${\rm Re}\,\Omega$ selected by the admissibility condition    (\ref{jump1}, \ref{A1})  $\Omega(X)|_{X\to \Gamma}>0 $.
 ${\rm Re}\,\Omega$  vanishes at
$
{\rm Im}\,\big(XY\big)=2tp(X).
$
There are a total of seven anti-Stokes lines connected at  branch points. They are transcendental and computed numerically, 
FIG.~\ref{beforeafter}.

Among the seven anti-Stokes lines, only three lines  $\Gamma_3, \Gamma_2$ and $\Gamma_1$ obey the second condition (\ref{A1}).  Line $\Gamma_1$ is the boundary of the finger. Lines $\Gamma_3, \Gamma_2$ are shocks. 

Selection works as follows. A  remote part of the finger  (${\rm arg \,X}=\pi,\;|X|\to\infty$) is not affected by the transition.  This selects a branch of $\Omega\sim\frac45X^{5/2}$ such that on the upper side of  $\Gamma_1$, ${\rm Re}\,\Omega>0$.
and therefore  in both sectors $\{{\rm IV, III}\}$ where  $|{\rm arg}\,X-\pi|<2\pi/5$ at a large $X$.  
Then signs of ${\rm Re}\,\Omega$ 
 are opposite on both sides of $\Gamma_4,\Gamma_5,\Gamma_6,,\Gamma_7$, and signs are positive on both sides of $\Gamma_1,\Gamma_2,\Gamma_3$, as required  by  \eqref{A1}. 

The Rankine-Hugoniot conditions (\ref{rh1} and (\ref{jump1}) give the velocity of shocks. Noticing that $\mbox{Im}\,XY=\sigma X_\perp$, where $X_\perp$ is a projection of a vector-coordinate of  a shock to a direction normal to  the shock, we get
$
V_\perp=\frac{X_\perp}{t}.
$

Already scaling yields that shocks push the fluid away faster ($\sim\! t^{5/4}$) than it is absorbed by the drain $\sim\! t$. {Therefore,  the finger retreats $v(e_3)\!=\dot e_3\!<\!0$,  smoothing the tip. Indeed, at the endpoints  $\Omega(e_{1,2})$ is purely imaginary.  Then by virtue of (\ref{jump1}) $\mbox{Im}\,\   \Omega(e_{1,2})=\int^{e_{1,2}}_{e_3}\sigma \dd\ell$ is a  mass deficit accumulated on shocks.  Eqs.(\ref{Omega},\ref{nn}) yield:
 \be
 \int^{e_{1,2}}_{e_3}\sigma \dd\ell=\frac 45 |\psi(e_{1,2})|= \frac45 (6.34513\dots)\,t^{5/4}\ .
 \ee
Since the boundary of the fluid moves towards lower pressure,  pressure is positive close to  sides of the tip, but remains negative around a distant part of the finger. Therefore, in the fluid there are
lines of zero pressure (the bright dotted line in FIG. \ref{beforeafter}). They emanate from the end points   crossing $\Gamma_1$ normally at  a  point  $x\!=\!-\eta t^{1/2}$, where velocity  vanishes,  FIG.~\ref{beforeafter}.   

Finally, we list  angles of zero-pressure line, the shock line and velocity at  $e_1$. Relative to the real axis, they are respectively $0.235061\pi,\, 0.3792582\pi,\,$ and $ 0.451357\pi$.  At $e_3$ the angle between shocks is  $2\pi/3$. 

\begin{figure}
\begin{center}
\includegraphics[width=0.25\textwidth]{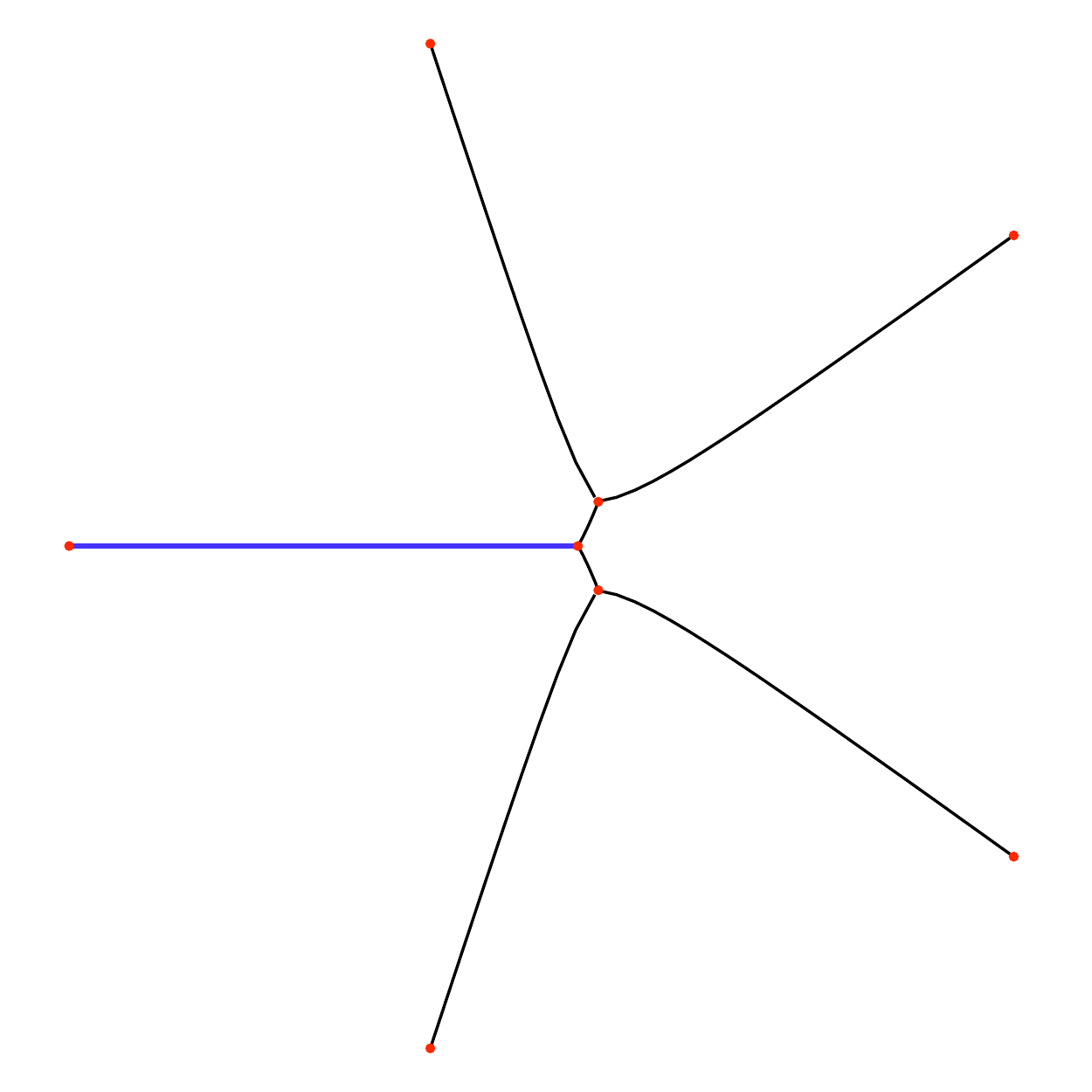}
\caption{A numerically computed graph of viscous shocks with two generations of  branching.}
\label{5legs}
\end{center}
\end{figure}

\paragraph*{10. Branching tree} Evolution through a (2,3)-cusp gives rise to a shock tree. It grows and  keeps branching further.  An interesting branching tree emerges, FIG.~\ref{5legs}. We do not know its global structure, but we do know that every generic branching event  is locally  identical to the transition we just described.  It will be  interesting to study whether a developed shock's graph with a large number  of branches exhibits a  scale invariance. 

\paragraph*{11.} Viscous shocks may have  different  meanings depending on  experimental settings. In a Hele-Shaw cell, viscous shocks are narrow channels where an inviscid liquid is  compressed and sheared. In a visco-ellastic fluid and in granular materials viscous shocks are cracks, etc. It will be very interesting to see an experimental realization of the branching tree of shocks in viscous flows at small Reynolds numbers.

\paragraph*{Acknowledgements}
S.-Y L. was supported by CRM-ISM postdoctoral fellowship.  P. W. was supported by NSF DMR-0540811/FAS 5-2783, NSF DMR-0906427, MRSEC under DMR-0820054 and the FASI of the Russian Federation under contract 02.740.11.5029.  P. W. and R. T. also acknowledge the USF College of Engineering Interdisciplinary Scholarship Program and the support of USF College of Arts and Sciences.  We thank A. Its and A. Kapaev for helpful discussions and are  especially grateful to I. Krichever.

\def\cprime{$'$} \def\cprime{$'$} \def\cprime{$'$} \def\cprime{$'$}
  \def\cprime{$'$}


\begin{thebibliography}{10}

\bibitem{Lamb}
H.~Lamb,
\newblock {\em Hydrodynamics}.
\newblock Cambridge Univ. Press, 1993.

\bibitem{Couder}
Y.~Couder,
\newblock {\em Perspectives in Fluid Dynamics}.
\newblock Cambridge Univ. Press, 2000.

\bibitem{DLA81}
T.  Witten,  L.M. Sander, 
\newblock {{\em Phys. Rev. Lett.}}, 47
:1400, 1981.

\bibitem{Jag-Nag}
X.~{Cheng} et. al.,
\newblock {\em Nature Physics}, 4:234
, 2008.

\bibitem{Maher}
H. Zhao and J.V. Maher,
\newblock {\em Phys. Rev. E}, 47
:4278, 1993.

\bibitem{Lipson}
S.~G. Lipson
\newblock {\em Physica Scripta}, T67:63--66, 1996.

\bibitem{ST}
P.  Saffman, G. Taylor,
\newblock {\em Proc. R. Soc. A}, \!245:312, 1958.

\bibitem{bs84}
B. Shraiman, D. Bensimon, {\em Phys. Rev. A}, 30:2840, 1984.

\bibitem{Howison85}
S.D. Howison, J.R. Ockendon, and A.A. Lacey,
\newblock {\em Quart. J. Mech. Appl. Math.}, 38
:343
, 1985.
\bibitem{Hohlov-Howison94}
Y.E. Hohlov and S.D. Howison, {\em ibid}, 51(4):777, \!1993.

\bibitem{Galin}
L.A. Galin,
\newblock {\em Dokl. Acad.Nauk  SSSR}, 47
:250, 1945.

\bibitem{PK}
P.~Ya. Polubarinova-Kochina,
\newblock {\em ibid.}, 47:254
, 1945.

\bibitem{Richardson72}
S.~Richardson,
\newblock {\em J. Fluid Mech.}, 56:609
 1972.

\bibitem{us1}
M. Mineev-Weinstein, P.B. Wiegmann, and A. Zabrodin,
\newblock {\em Phys. Rev. Lett.}, 84:5106, 2000.

\bibitem{us3}
I.~Krichever et. al.
\newblock {\em Physica D}, 198:1, 2004.

\bibitem{us4}
R. Teodorescu, A. Zabrodin, and P. Wiegmann,
\newblock {\em Phys. Rev. Lett.}, 95:044502, 2005.

\bibitem{us5}
E. Bettelheim et. al.
\newblock {\em Phys. Rev. Lett.}, 95:244504, 2005.

\bibitem{us6}
S-Y. Lee, E.~Bettelheim, and P.~Wiegmann,
\newblock {\em Physica D}, 219:22, 2006.



\bibitem{Howison86}
S.~D. Howison,
\newblock {\em SIAM J. Appl. Math.}, 46
:20--26, 1986.


\bibitem{Evans}
L.C. Evans, 
\newblock {\em Partial differential equations}, 
\newblock AMS, 1998.

\bibitem{LL} L. D. Landau and E. M. Lifshits. Fluid Mechanics, Butterworth-Heinemann, 1987.


\bibitem{Krichev-red}
I.M. Krichever, 
\newblock {\em Comm. \!Pure\! Appl.\! Math.}, 47
:437, 1994.

\bibitem{Bertola-Mo}
M. Bertola and M.Y. Mo,
\newblock {\em [math-ph/0605043]}, 2006.

\bibitem{Gamba} A genus transition of an elliptic 
curve is related to a semiclassical description of a 
solution of Painlev\'e I equation of  
F.~Fucito et al., 
\newblock {\em Int. J. of Mod. Phys. B}, 6:2123, 1992.

\end{thebibliography}
\end{document}